\documentclass[iop,apj,natbib]{emulateapj}
\pdfoutput=1

\slugcomment{Published in Science in School 39 \& 40 (2017)}
\shorttitle{Simple demonstration experiments for astronomical parallax}
\shortauthors{M. P\"ossel}

\begin{document}

\begin{abstract}
Parallax is the most important astronomical method for determining the distances from the Earth to
nearby stars. We present two hands-on activities that explore the concepts and practice of parallax measurement. The first activity is a simple geometrical set-up using simple theodolites, and is accessible for students grade 7 and older. The second activity uses digital photography to recreate astronomical parallax measurements in a classroom setting, and is suitable for students grade 10 and older.
\end{abstract}

\title{Simple classroom experiments for astronomical parallax}
\author{Markus P\"ossel}
\affil{Haus der Astronomie and Max Planck Institute for Astronomy\\
K\"onigstuhl 17, 69117 Heidelberg}
\email{E-mail: poessel@hda-hd.de}
\maketitle

\section{Introduction}

Astronomical parallax measurements use simple geometry to deduce the distances of objects within our solar system as well as in our galactic neigbourhood. They constitute one of the first rungs of the {\em cosmic distance ladder:} astronomy's network of methods of distance measurements, each valid for a range of distances, methods for greater distances calibrated using those for nearer. \citep{Grijs2011, Webb1999}. 

The fundamental concepts of this distance ladder are part of the curriculum of introductory astronomy courses \citep{CobleEtAl2013}, and the concept of parallax measurements is routinely taught in that setting. In addition, a number of educational hands-on activities exist, which allow students to gain first-hand experience either with real parallax measurements or with the basic concept. Astronomical parallaxes have been measured successfully with smaller and larger telescopes as part of introductory laboratory exercises \citep{RatcliffEtAl1993} and as high-school level research projects \citep{CenadelliEtAl2009,PenselinEtAl2014,CenadelliEtAl2016}. 

This article instead focusses on two simple demonstration experiments suitable for teaching the basic concepts of parallax measurements. Variants of these experiments have been used in astronomy education for decades \citep{Ferguson1977,DeJong1972,Deutschman1977}.  The purpose of this article is to provide updated descriptions of two classroom activities that allow students to explore parallax measurements in a simplified setting. Edited versions of the two parts of this article have been published (separately) in {\em Science in School} \citep{Poessel2017a,Poessel2017b}. 

\section{Background}

Astronomers observe from afar, separated by great distances from the objects they explore and examine. In this situation, knowing the distances of objects is crucial. Basic astronomical observations measure only the apparent brightness of an object. Unless we can also determine that object's distance, we will not be able to tell how much light the object emits. We could not distinguish an object that emits less light, but is closer to us, from an object that emits copious amounts of light but is more distant. 

Parallax is a basic effect most directly demonstrated by stretching out one of your arms, giving a thumbs-up, and observing your thumb first with one, then with the other eye. Your change in perspective from one eye to the other changes the apparent position of your thumb in front of the distant background. Hold your thumb closer, and you will see that the change in position increases. The magnitude of the change in position -- your thumb's parallax -- can be used as a measure of your thumb's distance!

Historically, parallax measurements have been used to determine the distance to the Moon, as well as the astronomical unit, but their most important contribution to astronomy are the first determinations of stellar distances. The capability to tell the distances to the nearest stars, and to determine the amount of light they emit, was a key step towards understanding the stars' physical properties, internal constitutions, and source of energy. 

The distances to the stars are so large that the accuracy needed for determining stellar parallaxes was beyond the reach of astronomers until the mid-19th century. Even so, the baseline --- the distance between the two different observer positions --- needed to be extremely large. Distances on Earth were significantly too small, and astronomers fell to making observations half a year apart: in that time, the Earth has moved along half its orbit, corresponding to a change in location of 2 astronomical units, or 300 million kilometers. 

The first widely accepted stellar parallax measurement was made by 
Friedrich Wilhelm \cite{Bessel1838}, soon followed by more precise measurements, with a jump in accuracy in the early 20th century due to the advent of astrophotography \citep{Hirshfeld2013}. 

The current gold standard is set by the ESA space mission Hipparcos, based on measurements taken between 1989 and 1993 and resulting in a catalogue of 100,000 stellar parallaxes with accuracies of less than a milli-arc second \citep{Perryman2010, vanLeeuwen2007}. It is due to be replaced by the results of the ongoing ESA mission Gaia, launched in late 2013. The mission's key goal is to determine position, proper motion and parallax for one billion stars with unprecedented accuracy, extending distance estimates with an accuracy of 10\% as far as the galactic center and reaching parallax accuracies of 20 microarcseconds for stars brighter than about 13 mag \citep{Prusti2012, deBruijne2014}.

\begin{figure}[htbp]
\begin{center}
\plotone{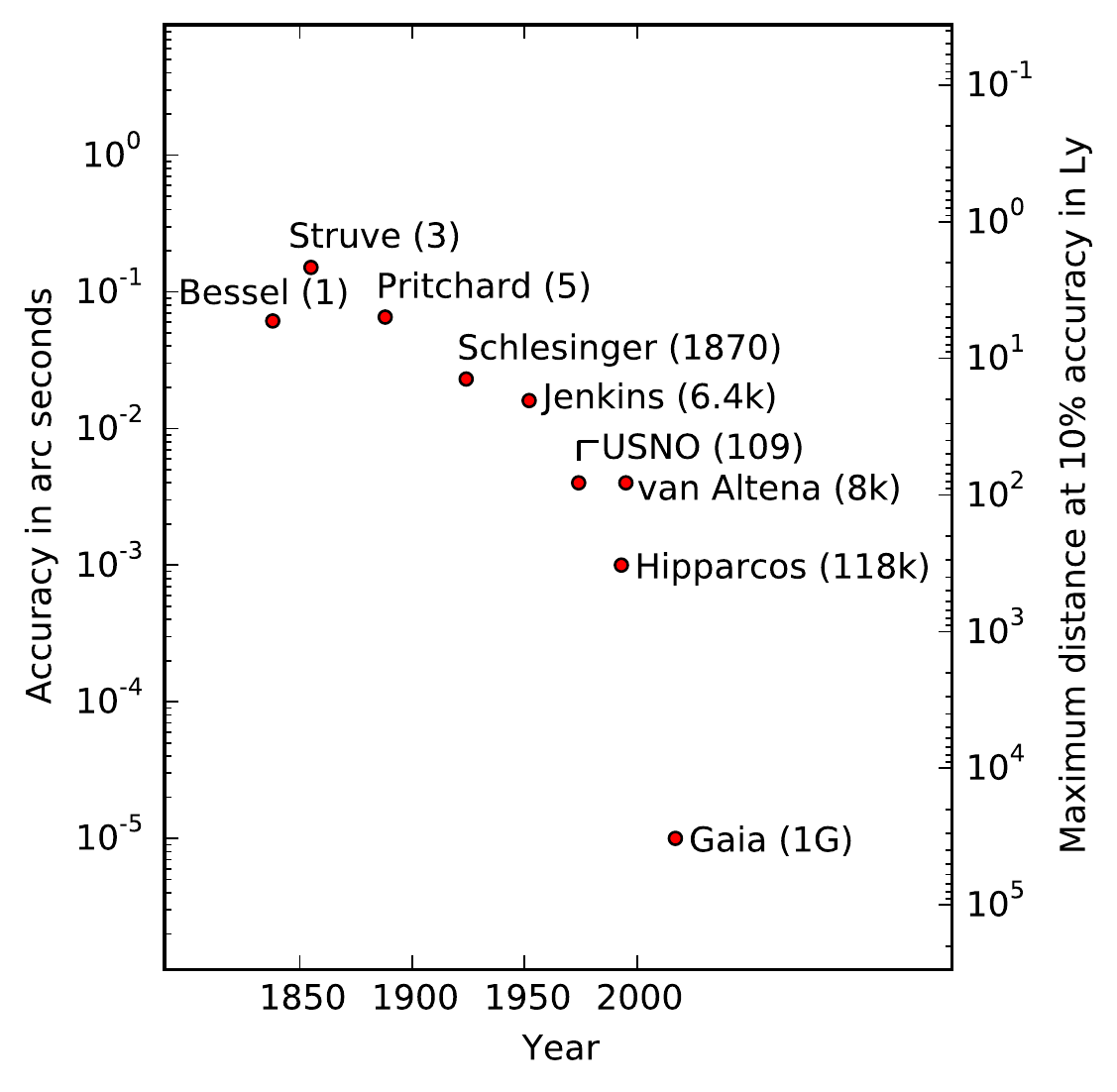}
\caption{Estimated parallax accuracy over time. Numbers in parentheses are the number of stars observed for a given survey.}
\label{ParallaxOverview}
\end{center}
\end{figure}

An overview of the evolution of the accuracy of parallax measurements over time is given in figure \ref{ParallaxOverview}.

\section{Parallax with simple theodolites}

The first experiment involves simple theodolites: instruments that can be used to measure the angles between two lines of sight. It is designed to illustrate a basic feature of all parallax measurements: how to measure distances to distant objects if all your measuring activities are confined within a comparatively small region of space.

To this end, the room in which the experiment is set up is symbolically divided into ``Earth'' and ``space''; measurements performed exclusively ``on Earth'' are used to reconstruct the distance to a model star located in ``space'', using simple trigonometric formulae or scale drawings. 

During internships at Haus der Astronomie, this experiment has been performed successfully by pupils between grades 7 and 11. It can also serve as a playful introduction for students who then go on to perform the photography experiment described in section \ref{ParallaxCamera}.

\begin{figure}[hbtp]
\begin{center}
\plotone{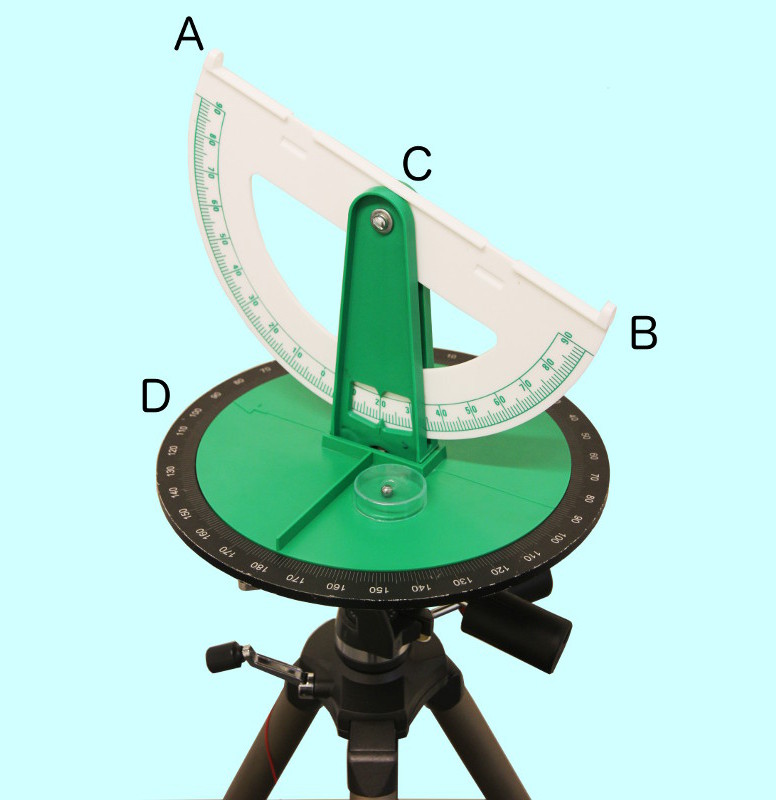}
\caption{Simple plastic theodolite for use in schools. The theodolite is aimed at the target by sighting the target along the aperture sights A and B. Using the turntable D and axis C, the theodolite can be aimed in any desired direction (except considerably downward). Once the theodolite is aimed at a target, azimuth and declination can be read off the built-in angular scales.
}
\label{TheoImage}
\end{center}
\end{figure}

Theodolites are instruments designed to measure a direction, described by a declination and an azimuth angle. Imagine the Earth as a transparent sphere; a theodolite placed in the center and oriented parallel to the equatorial plane could specify the direction to each location on the Earth's surface by measuring geographic latitude (declination) and longitude (azimuth angle). Precision theodolite are key instruments for surveying; a simple plastic model for use in schools is pictured in figure \ref{TheoImage}.

The turntable should be adjusted (e.g.\ using a tripod head) to be parallel to the ground. In our case, the turntable has a bull's eye spirit level attached for that purpose. The experiment only uses the theodolites' azimuthal degree of freedom only; the declination will be fixed at $0^{\circ}$ so that the sighting device is horizontal. 

With this restriction, it is possible for schools that do not own and cannot or do not want to purchase simple theodolites to build a simple replacement: a flat wooden board with a printed-out circle, marked from 0 to $360^{\circ}$, as the base for the turntable, another board as the turntable with a simple sighting device (e.g. two nails) attached can perform the basic measurements needed for this activity. 

\begin{figure}[htbp]
\begin{center}
\plotone{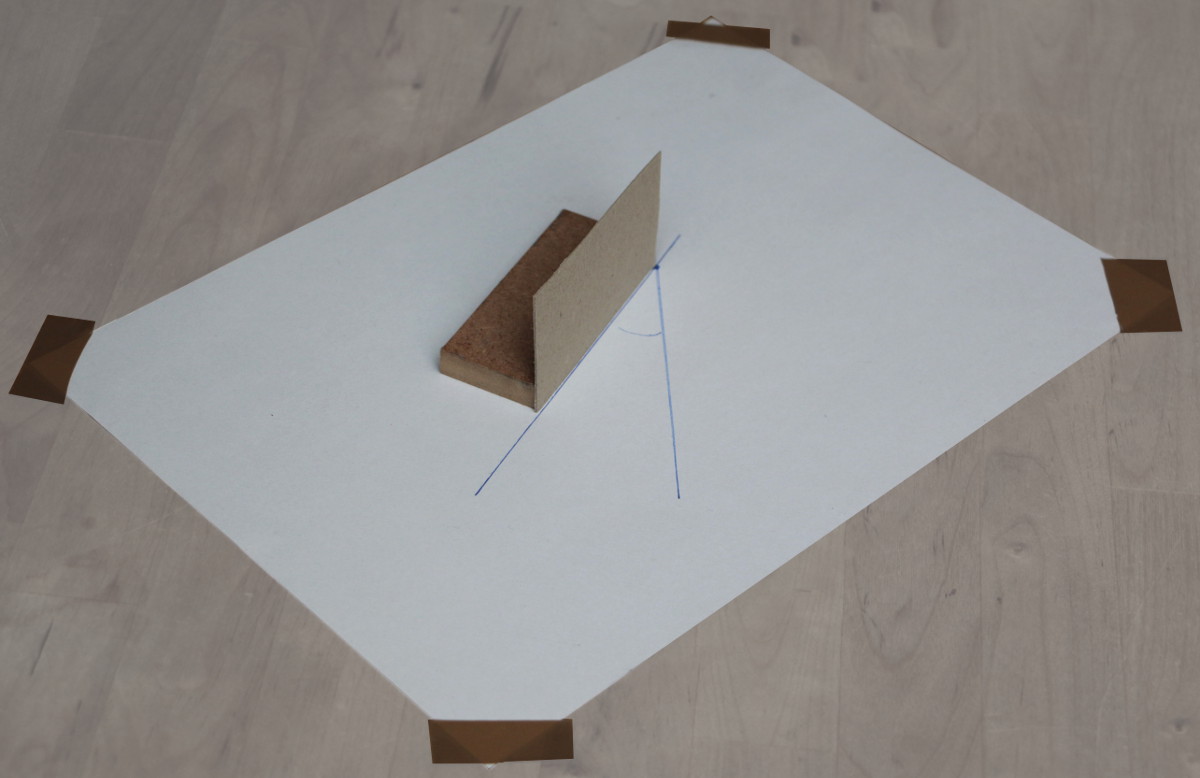}
\caption{Improvised instrument for measuring azimuth angles.}
\label{BlockReplacement}
\end{center}
\end{figure}

An even more simple replacement is shown in figure \ref{BlockReplacement}.
In this case, the sight is a piece of cardboard stabilised using a small wooden block. The block is placed on a piece of paper that has been affixed to a plane surface. The observation position is marked on the paper; to document the direction from this position to a distant object, aim the cardboard sight so that you look at the cardboard from the side, exactly edge-on, with the target lined up directly behind that thin cardboard edge. Make sure the bottom edge of the cardboard touches the point marking the observation position. Document the orientation of the cardboard edge by drawing an appropriate line onto the paper. Repeat for second target object; relative azimuth angles can be measured using their sight lines on the paper and a protractor. This simplified measurement device is similar to the one described in \citep{Ferguson1977}.

On to the rest of the setup: Our model star is a plastic sphere 15 cm in diameter. Other possibilities include a small LED, which would introduce an additional small measure of realism.

\begin{figure*}[htbp]
\plotone{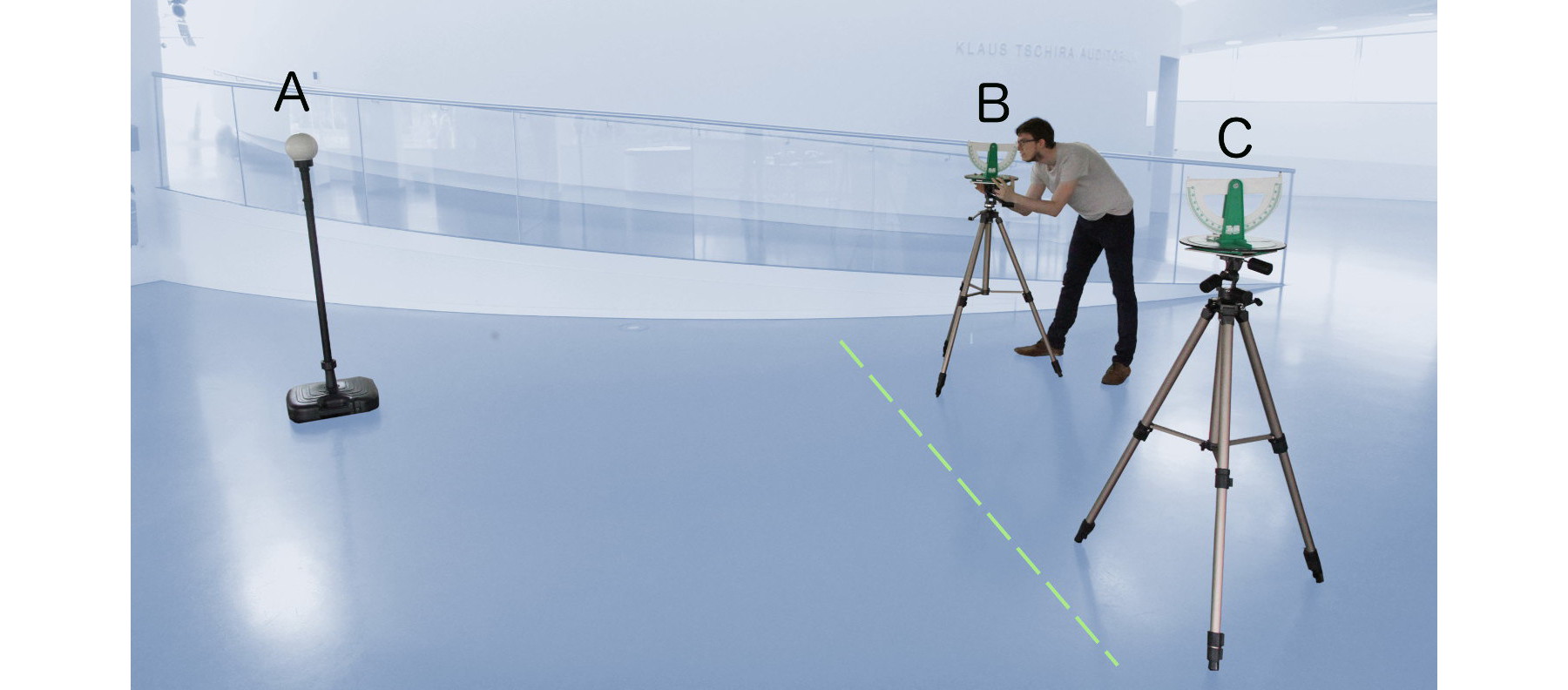}
\caption{Setup for the theodolite experiment: The dashed line separates ``Earth'' (right) from ``space'' (left). The two theodolites B, C on Earth are used to measure the distance to the star A in space by reconstructing the triangle $\triangle$ABC.
\label{TheoSetup}}
\end{figure*}
The basic set-up can be seen in figure \ref{TheoSetup}: Two theodolites mounted on tripods are placed in the zone designated as ``Earth'', some distance $b$ apart. If necessary, the activity can also be completed with a single theodolite, which then needs to be moved to positions B and C one after the other; its successive positions should be marked e.g. with tape. 

The aim is to determine the distance from each theodolite to the model star, which is placed in the ``space'' zone, without leaving Earth. 

The triangle $\triangle ABC$ is defined by the star and the two theodolites, more precisely: by the center of the ball that represents the stars, and by the intersections of the theodolite's sight lines with their vertical axes.

Evidently, only three quantities of the triangle $\triangle ABC$ in figure \ref{TheoSetup} can be measured without leaving Earth: the baseline length $b\equiv \overline{BC}$ and the angles $\angle ABC$ and $\angle ACB$. Fortunately, by basic trigonometry, those three quantities suffice to define the triangle completely: The distances from each theodolite to the star are given via the law of sines by
\begin{eqnarray}
\label{TheoDist1}
 \overline{AC} &=&  \frac{\sin(\angle ABC)}{\sin(180^{\circ} - \angle ABC - \angle ACB)}\cdot b\\[0.5em]
\nonumber &\mbox{and}& \\[1em]
\label{TheoDist2}
\overline{AB} &=&  \frac{\sin(\angle ACB)}{\sin(180^{\circ} - \angle ABC - \angle ACB)}\cdot b,
\end{eqnarray}
respectively.

In preparation of the activity, the teacher has placed both the stars and the theodolite,  as well as marked a line (e.g. with a rope) separating Earth and space. The sighting devices of the theodolites and the center of the model star should be at the same height above ground. That way, the trigonometric problem is defined within a single plane, which is parallel to the ground.

Depending on the aims of the teacher, the activity itself can either be preceded by an introduction to parallax measurements or commenced by posing the central questions: Given the two theodolites and a tape measure, how can you determine the distance to the model star without leaving Earth?

A possible solution is as follows. First, measure the distance $b\equiv\overline{BC}$ using the tape measure. This is best done by projecting the points B and C to the ground using a plumb bob, and then measuring the distance between the projected points on the ground. (A direct measurement in the air, keeping the tape measure tight by pulling on both ends, is more cumbersome and less precise.)

To measure the angles, begin by aiming the two theodolites at each other in what we shall call the zero position. For each theodolite, determine the azimuth of the zero position. (In our setup, the base plate of the turntable can be moved so as to move the $0^{\circ}$ mark to this position.) Next, aim each theodolite at the star. The difference between the star's azimuth value and the zero position value yields the angles $\angle ACB$ and $\angle ABC$, respectively.

When sighting along a theodolite by lining up the instrument's two sights, keep some distance between your eye and the instrument. This will make it easier for your eye to keep both of the sights in focus at the same time. 

As for reading off angles, our theodolite marks the azimuth angle in full degrees. For our own replication of the activity, we divided the space between two degree markers (ticks) in four: lower degree marker $+0.5$ if the pointer appeared to be directly in the middle between the lower and higher degree marker, $+0.3$ if between the lower marker and the middle, and $+0.7$ if between the middle and the higher marker. 

Once these measurements have been made, older pupils can use the law of sines, formulae (\ref{TheoDist1}) and (\ref{TheoDist2}), to calculate the distances $\overline{AB}$ and $\overline{AC}$. Younger pupils may draw the triangle, to scale, on a piece of paper; we have achieved fairly accurate results using a scale of 1:50 (that is, one meter drawn as two centimeters) on an A3 piece of paper ($29.7\,\mbox{cm}\times 42\,\mbox{cm}$).

After the calculations have been made, the pupils can check their results by ``flying into space,'' that is, by measuring  $\overline{AB}$ and $\overline{AC}$ directly using a tape measure. 

The activity can be repeated for various distances to the model star, or for multiple stars placed at various distances. 

A variation worth thinking about is to measure angles not relative to the other theodolite, but relative to a very distant feature of constant location. 
This requires free sight into the distance, and makes the activity a bit more complex, but it does bring matters closer to how astronomers actually measure parallax, namely relative to much more distant reference objects (cf. section \ref{ParallaxCamera}).

\begin{figure}[htbp]
\begin{center}
\plotone{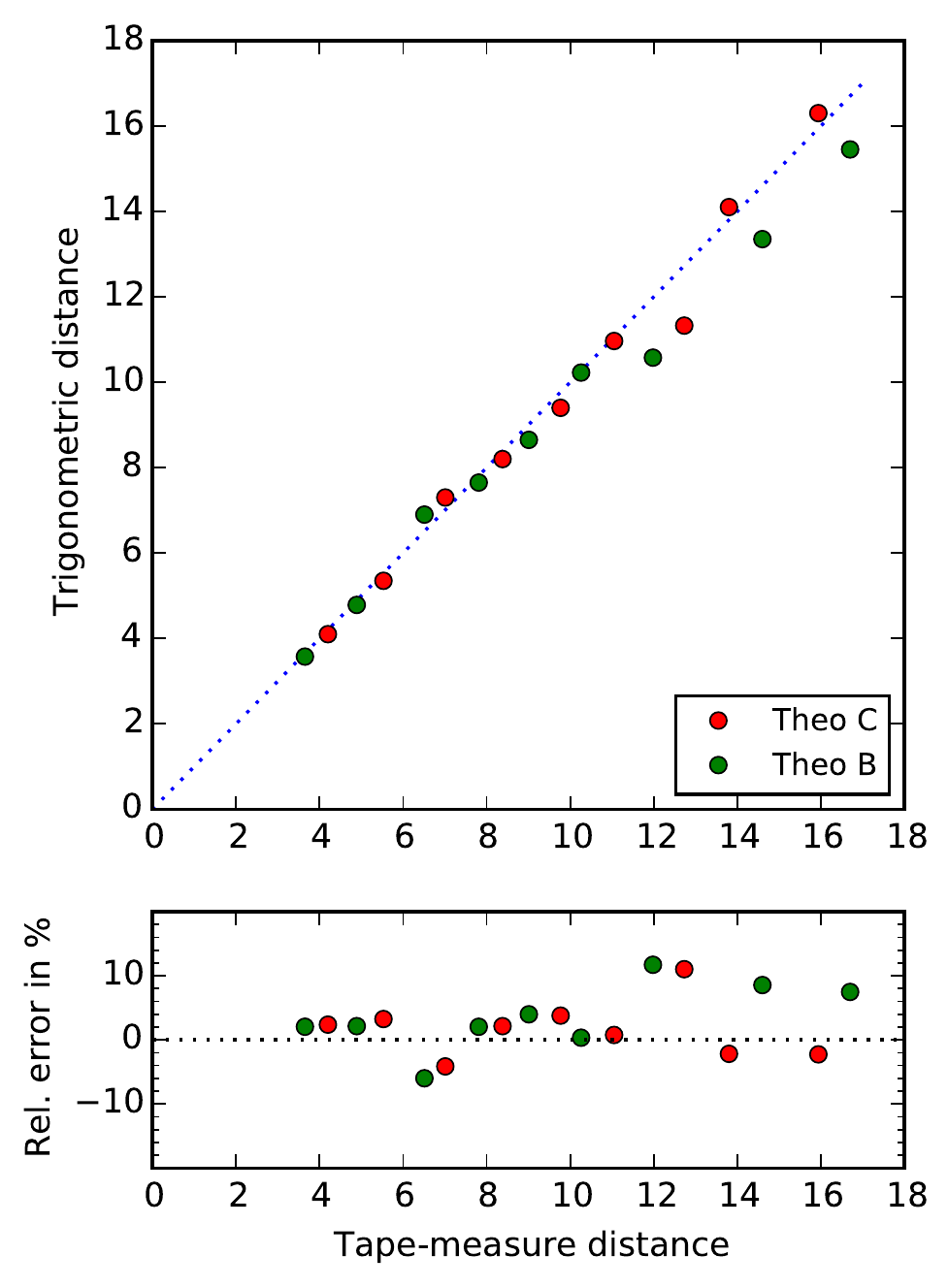}
\caption{Top: tape-measure distances (x axis) vs. parallax distances (y axis). Parallax distances have been reconstructed by drawing triangles to scale. Red dots represent the distance between the theodolite at C and the model star, green dots between theodolite B and the model star. Perfect correspondence between trigonometric and tape measure distance values would place all data points would lie on the dotted line. Bottom: relative error, defined as the deviation of tape-measure and trigonometric distance relative to the tape-measure distance.
}
\label{MeasurementsTheodolite}
\end{center}
\end{figure}

For interstellar distances up to about $16\,\mbox{m}$, we have found that distance measurements within about $10\%$ are readily obtainable. The accuracy of sample measurements we made specifically for this article is shown in figure \ref{MeasurementsTheodolite}, and all errors are below $14\%$. The dominant error source are the angular measurements; in particular, analysis via scale drawing does not increase the errors significantly. When we had different groups of pupils working on this activity, mean accuracy ranged between $10\%$ and about $50\%$ (reconstruction via scale drawing).

After the pupils have completed the activity, it makes sense to have them figure out the answers to questions about their measurements and about the link between their measurements and astronomy: How large was the parallax angle $\angle BAC$ in their case? How accurate do they think they were able to read off the azimuth angles? (With our read-out recipe, the answer would be: about $0.3^{\circ} =  18'$.) With the chosen baseline, what distances can they determine with an accuracy of, say, $10\%$? How does this compare to astronomical distances? What can be done to improve accuracy? (Increase precision of instruments, increase the baseline.) What is the largest baseline on Earth? (Twice the Earth's radius.) 

What distances can be measured with this baseline and a read-off accuracy as in the activity? With a read-off accuracy of $0.1''$? With an accuracy on the order of 1 milli-arc second like Hipparcos, or 10 micro-arc seconds like Gaia? Once even those accuracies are seen to be insufficient, it's time to bring in a much larger baseline: the Earth's orbit around the Sun, where observations timed half a year apart will yield a base line of 2 astronomical units, or 300 million kilometers. This brings distances on the order of light years and more into reach. 

In our experience, pupils are likely to need some guidance in order to find this solution for the longer baseline, although some might reach this conclusion on their own.

\section{Parallax with a digital camera}
\label{ParallaxCamera}

Closer to real astronomical parallax measurements both in concept and in practice is the following activity, which we have undertaken successfully with pupils grade 10 and older: use two images taken with a digital camera from two different locations to determine the distance to an object that is visible in both images.  Analog photography versions of this activity have been in use in undergraduate astronomy courses \citep{DeJong1972,Deutschman1977}. The digital version described here requires no more than the ability to read off image pixels in an image manipulation program such as Adobe Photoshop or The Gimp, as well as knowledge of the rules governing similarity in triangles. 

\begin{figure}[htbp]
\begin{center}
\plotone{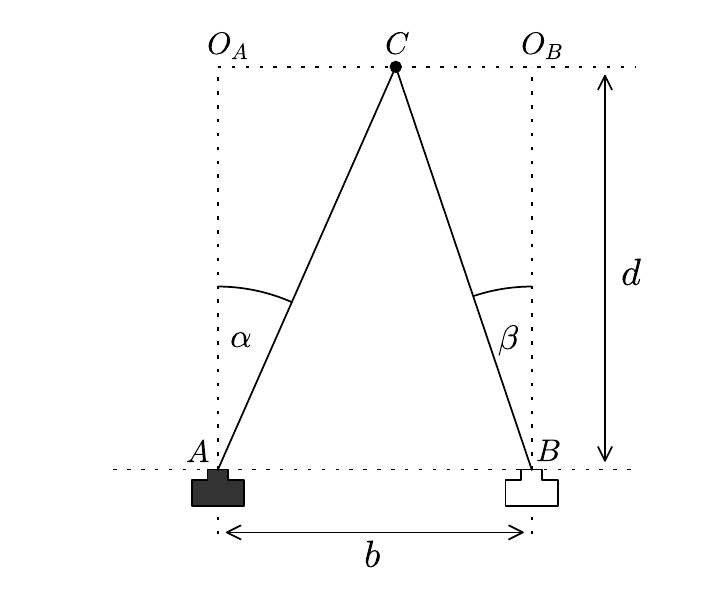}
\caption{Birds-eye view of the basic setup for the camera activity
}
\label{CameraSetup}
\end{center}
\end{figure}

The basic setup is shown in figure \ref{CameraSetup}: A digital camera is fixed first in location $A$, then in location B, taking a digital photo in each location. The target, the model star $C$, should be placed in the approximate viewing direction of the cameras, not off to the side. 

For simplicity, let us assume that both images are taken with the viewing direction of the camera orthogonal to the line $AB$. 

The two camera positions represent two positions of Earth, half a year apart, from which we observe the star $C$. Evidently, we have three pieces of information regarding the triangle $\triangle ABC$: the angles $\alpha$ and $\beta$ and the shift distance (baseline distance) $b$. This is sufficient to define the triangle, and in particular to fix its altitude $d$, the desired distance of the star $C$ from us.

At first glance, the trigonometry looks somewhat complicated, demanding the use of the law of sines and other identities. But in fact, the situation simplifies once we model the camera as a pinhole camera, and once we realise that there is an aspect of relativity in the situation: where figure \ref{CameraSetup} shows the star $C$ in a constant position, and the camera shifting position by a distance $b$ from $A$ to $B$ to the right, an observer sitting on the camera will see the star shift a distance $b$ in parallel with the line $AB$. 

\begin{figure}[htbp]
\begin{center}
\plotone{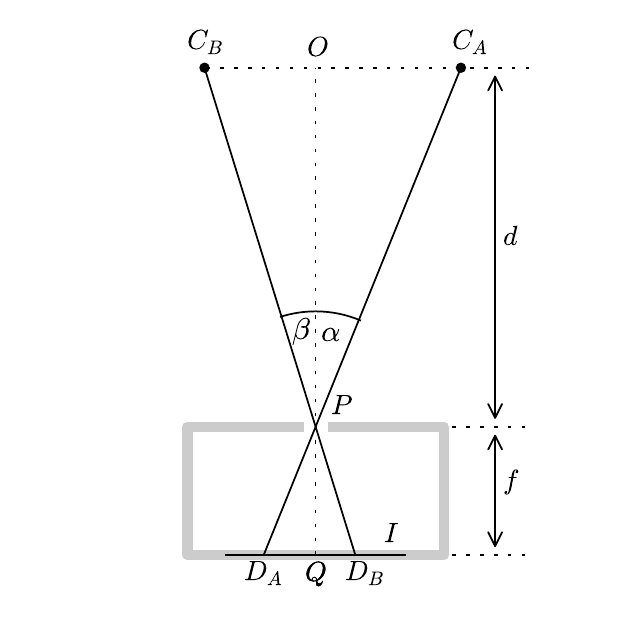}
\caption{Modelling the measurements with a pinhole camera 
}
\label{PinholeSetup}
\end{center}
\end{figure}

Once we combine these two elements, we obtain figure \ref{PinholeSetup}, showing a plane that intersects the camera and contains the star's apparent positions as well as the direction of shift. In this figure, $C_A$ is the apparent position of the star as seen by the camera when it was in position $A$, and $C_B$ is the apparent position of the star seen from the camera at position $B$. We have replaced the symbolic representation of the camera by a pinhole camera: light can only fall into the camera via the pinhole $P$. Once a ray of light enters, it continues straight on until it encounters the image detector plane $I$, whose center is marked as $Q$. Along the line joining $C_A$ and $C_B$, those two points are a distance $a$ and $b$ from the center $Q$ that marks the direction in which the camera is aimed. The rays of light coming from $C_A$ and $C_B$ and passing through the pinhole intersect the image detector plane at $D_A$ and $D_B$, respectively.

The key to a geometrical description are similar triangles. $\triangle C_BPO$ is similar to $\triangle D_BPQ$, since by definition these triangles share two angles. By the same reasoning,  $\triangle C_APO$ is similar to $\triangle D_APQ$. But similarity means that the ratios of corresponding sides of the triangle are equal:
\begin{equation}
\label{SimilarPinhole}
\frac{\overline{D_BQ}}{f} = \frac{\overline{C_BO}}{d}  \;\;\;\mbox{and}\;\;\; \frac{\overline{D_AQ}}{f} = \frac{\overline{C_AO}}{d}. 
\end{equation}
The relativity that brought us from figure \ref{CameraSetup} to figure \ref{PinholeSetup} is encoded in the facts that
$\triangle CBO_B$ and $\triangle C_BOP$ are congruent, and so are $\triangle CAO_A$ and $\triangle C_AOP$, and 
\begin{equation}
\overline{C_BO} + \overline{OC_A} = \overline{C_BC_A} = b.
\end{equation}
Using this and (\ref{SimilarPinhole}), we find the fundamental imaging equation for our setup, namely
\begin{equation}
d = \frac{fb}{\Delta D} 
\end{equation}
where $\Delta D\equiv \overline{D_AD_B}$. The image plane is where the detector is situated, dividing the plane into pixel segments. If the pixel scale is such that a distance $\Delta D$ in the image plane corresponds to a distance of $p = \lambda\,\Delta D$ pixels, we obtain
\begin{equation}
d = \frac{(\lambda f)\cdot b}{p} 
\end{equation}
as the relation between the star's distance $d$ and the pixel distance $p$ by which the star moves when the camera is shifted by the distance $b$. The proportionality factor $(\lambda f)$ must be determined by a calibration procedure. The simplest such procedure amounts to taking an image of an object of known length $L$ that is oriented orthogonally to the line-of-sight of the camera, at a known distance $d_L$. The object's length in pixels $p_L$ is read off the image. Then
\begin{equation}
(\lambda f) = \frac{p_L\cdot d_L}{L}.
\end{equation}

\begin{figure}[htbp]
\begin{center}
\plotone{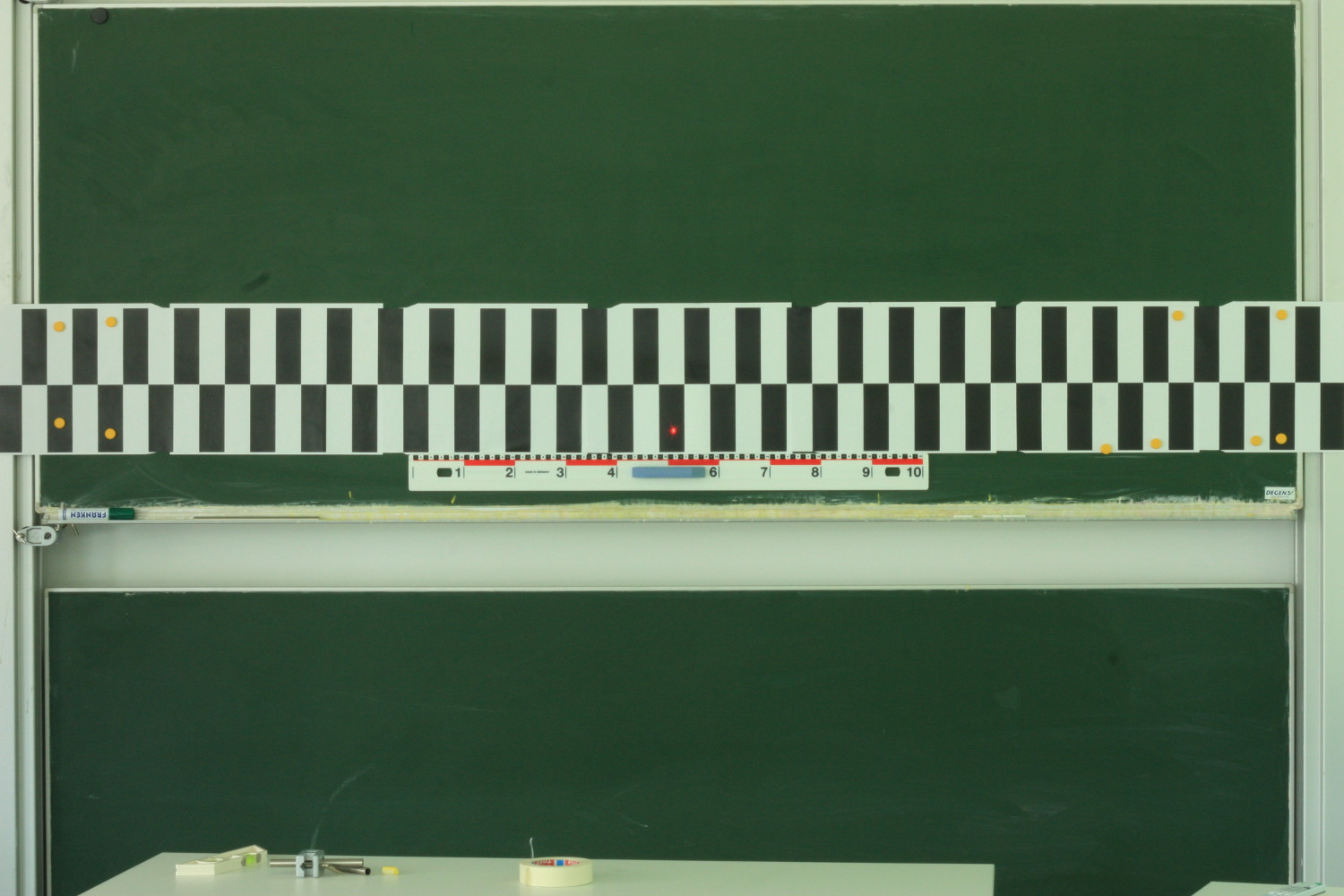}
\caption{Calibration pattern, taped horizontally to a blackboard, as seen by our camera from a distance of 6 m.}
\label{CalibrationPattern}
\end{center}
\end{figure}

For the activity, we used a Canon EOS 450D SLR camera (= Digital Rebel XSi) camera with the fixed focus length Canon EF  50 mm 1:1.8 II lens. At this focus length, approximating the camera's image map by that of a pinhole camera works reasonably well. For other camera--lens combinations, the teacher should test how well (or not) this approximation works before setting the pupils to work. In our case, we tested the model at the same time we performed the calibration: by taking an image of a pattern of alternating black and white rectangles (see figure \ref{CalibrationPattern}), each 5 centimeters in width, for a total length of 250 cm, taped horizontally across a blackboard, with the camera at a distance of $6.07\,\mbox{m}$ (as measured by a commercially available triangulation-based laser range finder, Bosch DLE 40, purchased at a DIY store). 

\begin{figure}[htbp]
\begin{center}
\plotone{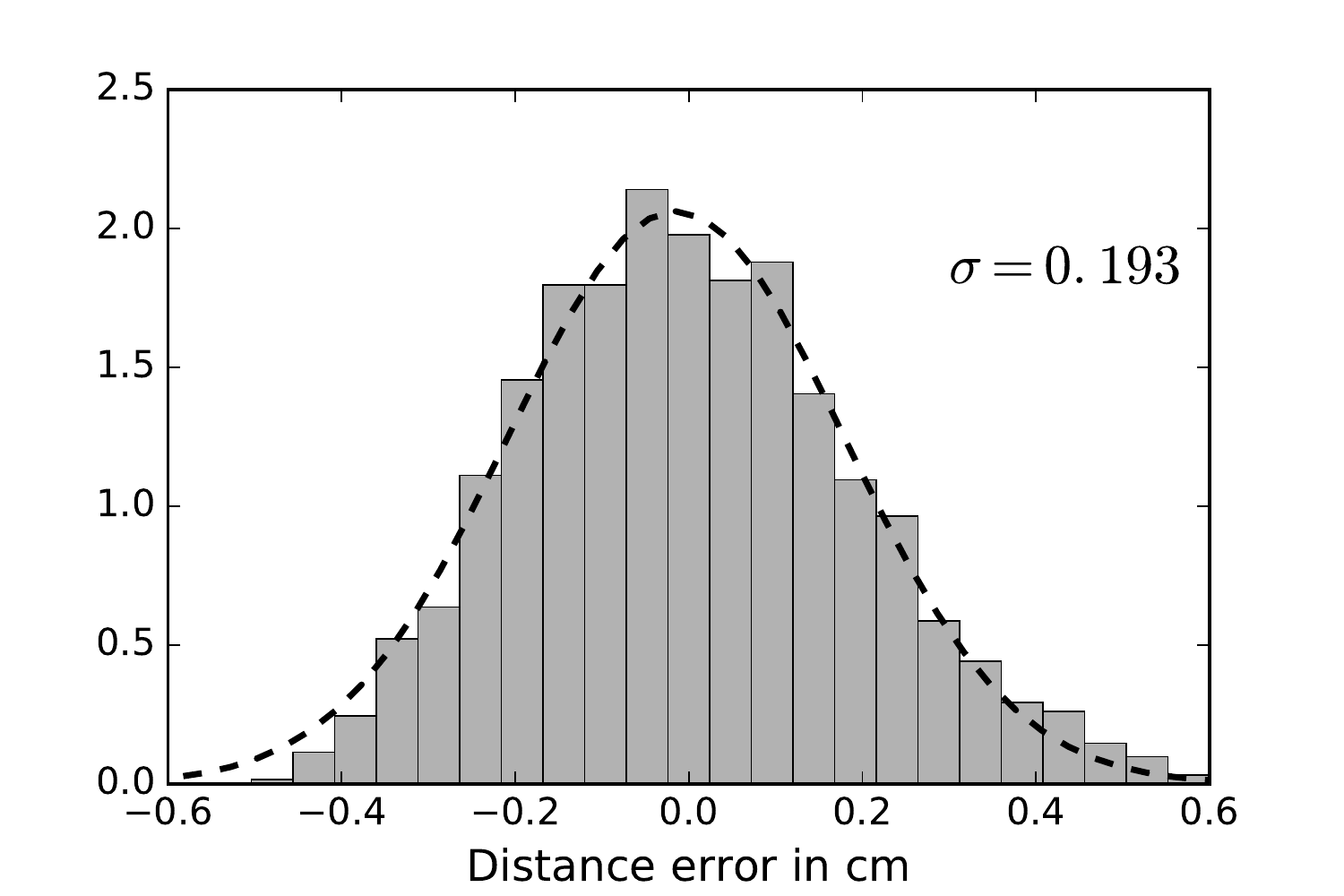}
\caption{Histogram of the differences between direct measurements of the distance between two vertical rectangle borders of the
calibration pattern and an estimate of such distances using the photographic image and the pinhole model.}
\label{ErrorHistogram}
\end{center}
\end{figure}

Fitting the distance of each vertical rectangle border from the center to the simple pinhole model, we obtained the calibration factor
\begin{equation}
\label{CalibrationFactor}
(\lambda f) = 9836\;\mbox{px}.
\end{equation}
This is close to what we would expect by equating $f$ and the nominal focus length of the lens used (50 mm), and then using the chip size given in the camera specifications, namely a row of 4272 pixels that corresponds to a chip width of 22.2 mm: For those values, one would get $\lambda f = 4272\cdot 50/22.2 = 9621\;\mbox{px}$. 

The adequacy of the pinhole model, and the quality of the calibration, can be estimated by comparing distances from the center of the pattern as measured on the pattern itself with distances determined using the camera image and the pinhole model with calibration factor (\ref{CalibrationFactor}). For all the possible comparisons between rectangle borders (distances from each border to all the different borders to the right of it), the maximum error amounts to 0.6 cm (corresponding to $3.4$ arc minutes). The histogram of the differences (residuals) can be seen in figure \ref{ErrorHistogram}. The standard deviation of a fitted Gaussian distribution is $\sigma = 0.193\,\mbox{cm}$ ($1.1$ arc minutes). 

\begin{figure}[htbp]
\begin{center}
\plotone{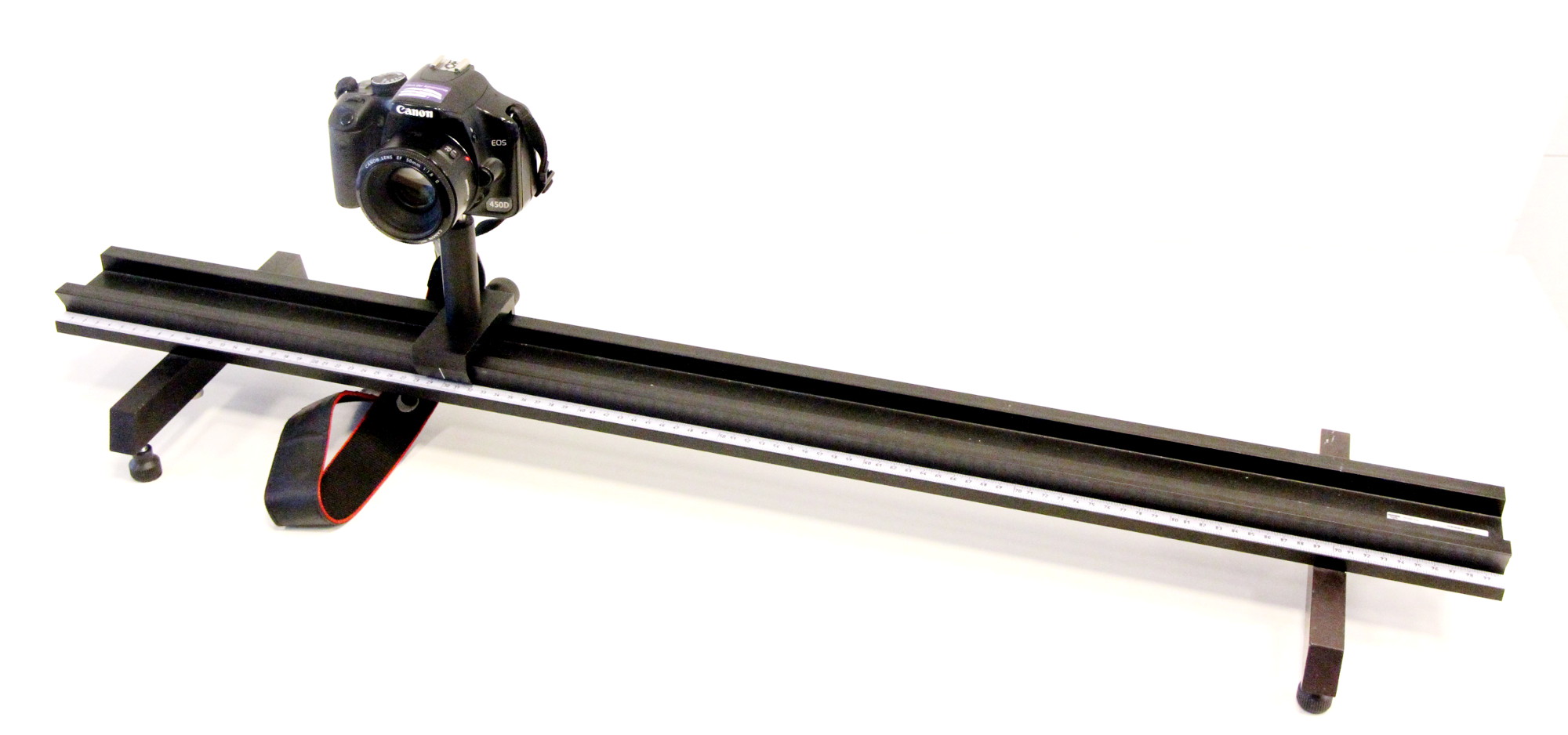}
\caption{Digital camera, mounted on, and pointing at right angles to, an optical bench. The white line on the front portion of the bench is a centimeter scale where the camera position can be read off.}
\label{CameraBench}
\end{center}
\end{figure}

For the activity itself, the camera was mounted on an optical bench for use in schools. The bench included a distance scale, allowing for easy determination of the shift distance (figure \ref{CameraBench}). The target representing the star, placed at distances between 2.5 and 7 m from the camera. For comparison purposes, these distances were read off a tape measure laid out straight on the set of tables that supported both the optical bench and the target, on which we shifted the target between measurements. 

In the simplified sketch in figure \ref{CameraSetup}, the camera is always pointing in the same direction. In practice, shifting the camera will sometimes result in small shifts of the camera's orientation. While one could mount the camera so as to avoid such shifts, it is educational not to take additional measures. After all, for real parallax measurements, astronomers cannot keep a telescope pointing in an unchanged direction for half a year, while Earth moves along its orbit from one observation point to the next. Instead, they will take an object that is considerably more distant than the target star; pointing the telescope at this object will ensure constancy of direction.

In our case, we chose a feature on one of the observatory domes of the Max Planck Institute for Astronomy as a marker indicating one particular direction, which was located at a distance of about 80 m from our camera. A more distant reference object would have been desirable, but was not available in our case. In measuring pixel shifts between one image and another, we aligned the images first, using the marker as a reference.

Pixel coordinates of our target object and our reference marker were determined by using the free and open source image processing software Gimp (\href{http://www.gimp.org}{http://www.gimp.org}). 

\begin{figure}[htbp]
\begin{center}
\plotone{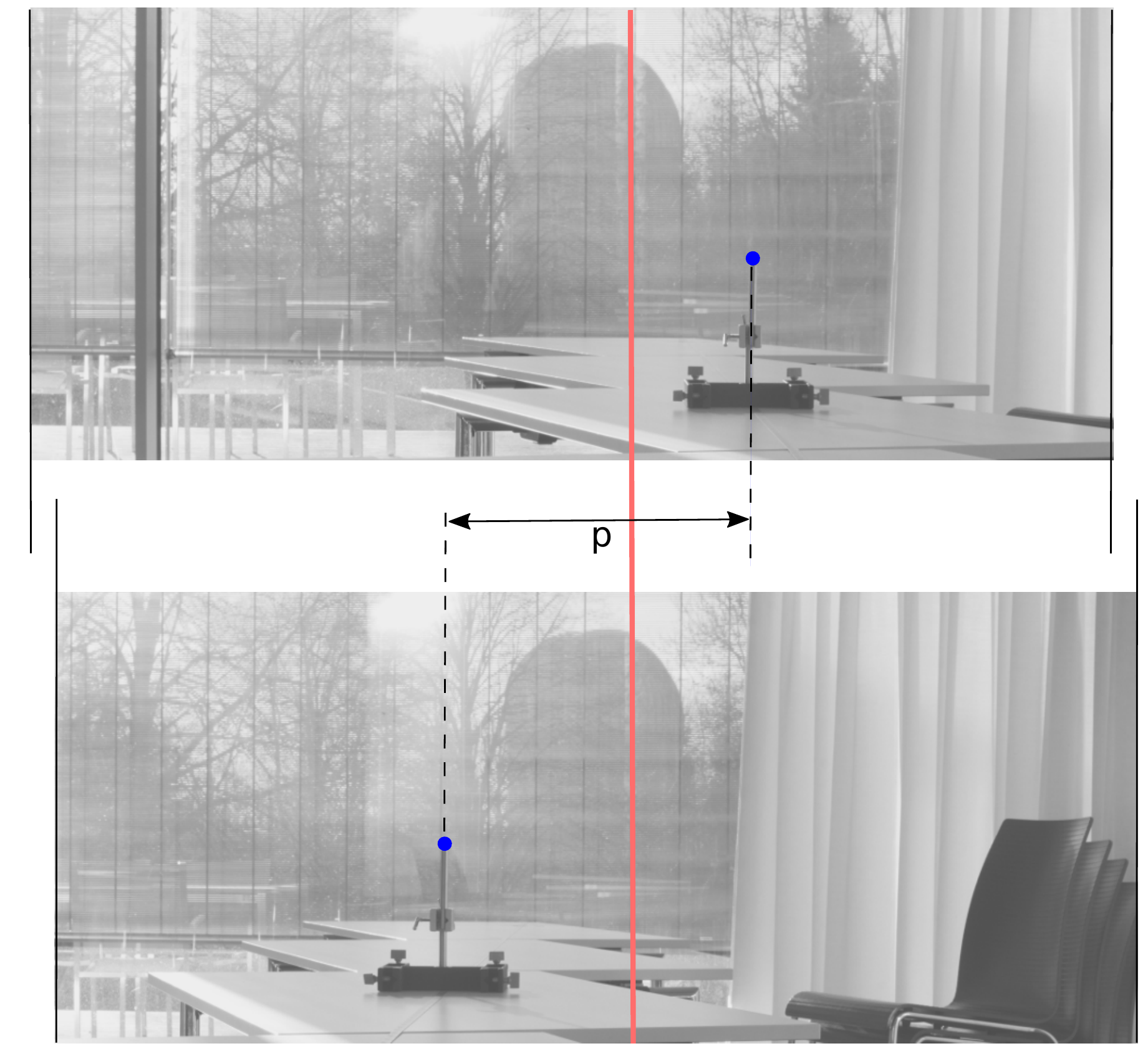}
\caption{Two sample images, simplified and cropped, from the camera activity, showing a horizontal pixel shift $p$. The images have been aligned using the position of a recognizable marker on the dome in the background; the images have been aligned horizontally so that in each image, that marker is on the red line.
}
\label{ImageShift}
\end{center}
\end{figure}

Two images from our activity, with the target at a distance of  4 m from the camera, can be seen in figure \ref{ImageShift} -- at least in simplified, cropped form. The top image was taken with the camera in what would be position $A$ in figure \ref{CameraSetup}, the lower image from position $B$. In each image, the model star is marked (slightly exaggerated in size) as a blue disk. The images have been aligned relative to the vertical red line using the aforementioned structural marker on the dome (which is too small to be discernible in these images). In comparing the different positions of the model star, we find a horizontal shift $p$ that encodes stellar parallax. 

With the parallax pixel shift $p$ from the (aligned) images, using the calibration factor (\ref{CalibrationFactor}) and the shift distance $b$ that was chosen for the camera, which amounts to 50 cm in our case, we can determine parallax distances for our model star. 

\begin{figure}[htbp]
\begin{center}
\plotone{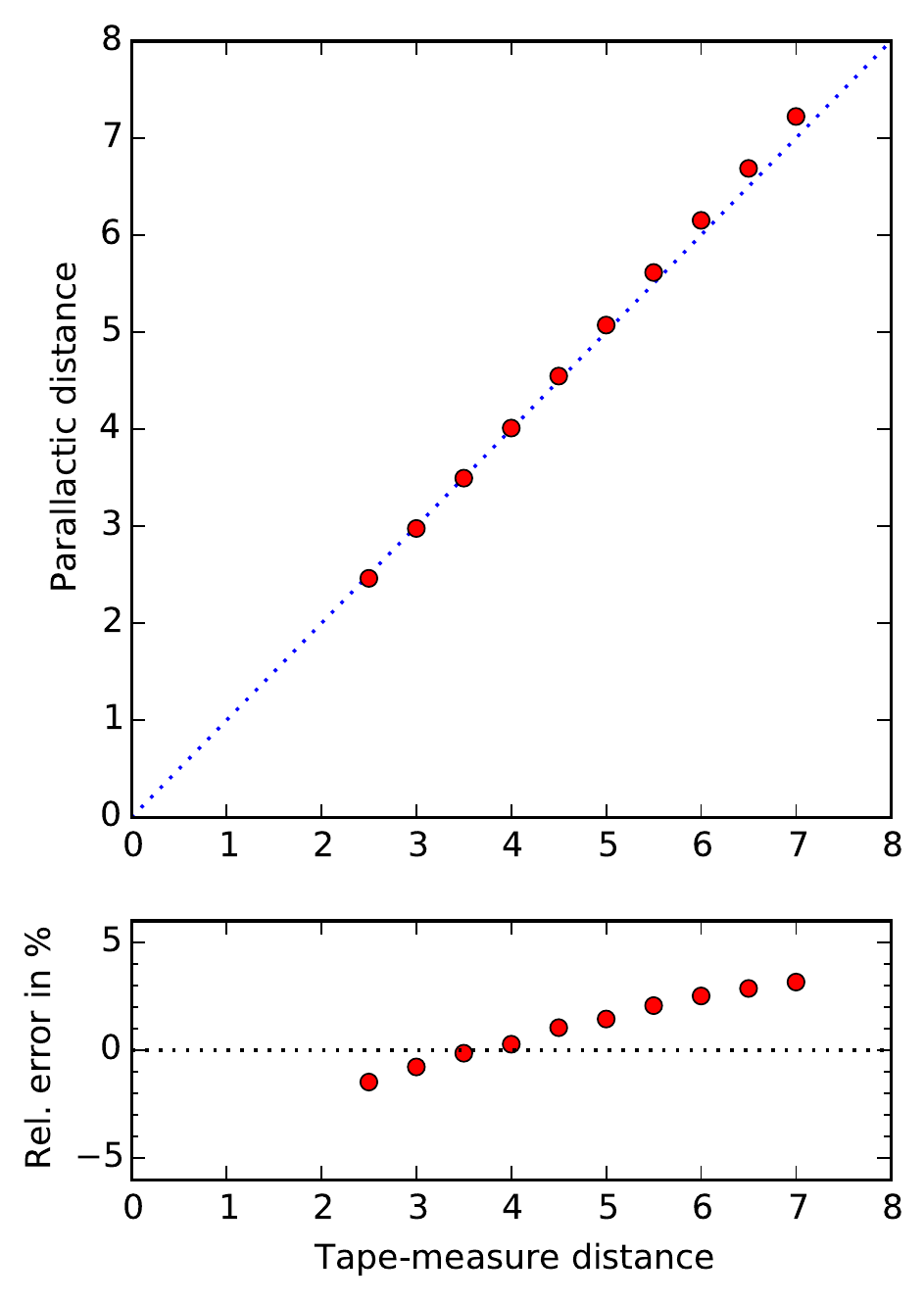}
\caption{Top: Distances to the model star measured by the parallax method plotted against tape-measure distances. Bottom: Relative errors of the parallax measurements, determined relative to the tape-measure distances, in percent. 
}
\label{CameraAccuracy}
\end{center}
\end{figure}

The results can be seen in figure \ref{CameraAccuracy}. Visual inspection confirms good agreement between the two different methods. The largest relative error, for the largest measured distances, amounts to $3.2\%$. 

Qualitatively, an increase of the relative error is expected for larger distances as the (scale-free) geometry changes: The distance to the model star becomes larger compared with the distance of the reference object; hence, the error introduced by the parallax of the reference object becomes ever greater. 

\section{Conclusion}

The activities described here provide demonstrations of the principle of parallactic distance measurements at different levels: with a focus on the basic geometry in the case of the activity using the simple theodolites, and closer to the way astronomers actually measure parallax for the camera activity. 

Both activities are hands-on, contrasting a tape-measure control measurement with the parallax method. The mathematics involved is such as not to discourage younger pupils.

Both activities provide pupils with a sense of the accuracy of the measurements involved: from around 10\% for the theodolites to 3\% (and expected to be better when a more distant reference object is used) for the camera activity.

Simple calculations involving ratios can serve to link these accuracies in a classroom setting to actual parallax measurements, allowing pupils to appreciate the precision reached by the ESA mission Hipparcos and that of the ongoing ESA mission Gaia.

\section{Acknowledgement}

I would like to thank Paul Eckartz for his help with figure \ref{TheoSetup}.


\begin{thebibliography}{natbib}
\bibitem[Bessel(1838)]{Bessel1838} Bessel, F.W.: ``Bestimmung der Entfernung des 61sten Sterns des Schwans'' in {\em Astronomische Nachrichten} {\bf 16} (1838), 65--96. doi: \href{http://dx.doi.org/10.1002/asna.18390160502}{10.1002/asna.18390160502}
\bibitem[de Bruijne et al.(2014)]{deBruijne2014} de Bruijne, J.H.J., K.L.J. Rygl \& T. Antoja 2014: ``Gaia Astrometric Science Performance --- Post-launch Predictions'' in N.A. Walton, F. Figueras, L. Balaguer-N\'u\~ez \& C. Soubiran (eds.): {\em The Milky Way Unravelled by Gaia: GREAT Science from the Gaia Data Releases}. EAS Publications Series, {\bf 67--68}, 23--29. doi: \href{http://dx.doi.org/10.1051/eas/1567004}{10.1051/eas/1567004}
\bibitem[Cenadelli et al.(2009)]{CenadelliEtAl2009} Cenadelli, D., M. Zeni, A. Bernagozzi, P. Calcidese, L. Ferreira, C. Hoang \& C. Rijsdijk 2009: ``An international parallax campaign to measure distance to the Moon and Mars'' in {\em European Journal of Physics} {\bf 30}, 35--46.
doi: \href{http://dx.doi.org/10.1088/0143-0807/30/1/004}{10.1088/0143-0807/30/1/004}
\bibitem[Cenadelli et al.(2016)]{CenadelliEtAl2016} Cenadelli, D.,  A. Carbognani, A. Bernagozzi \& C. Olivotto 2016: ``Geometry can take you to the Moon'' in {\em Science in School} {\bf 35}, 41--46. Online version: \href{http://www.scienceinschool.org/content/geometry-can-take-you-moon}{http://www.scienceinschool.org/content/geometry-can-take-you-moon}
\bibitem[Sec. III.C. in Coble et al.(2013)]{CobleEtAl2013} Coble, K. et al. 2013: \href{http://esoads.eso.org/abs/2013AEdRv..12a0102C}{``Investigating Student Ideas about Cosmology I: Distances and Structure''} in {\em Astronomy Education Review} {\bf 12}, 1. doi: \href{http://dx.doi.org/10.3847/AER2012038}{10.3847/AER2012038}
\bibitem[De Jong(1972)]{DeJong1972} De Jong, M.L. 1972: ``A Stellar Parallax Exercise for the Introductory Astronomy Course'' in
{\em American Journal of Physics} {\bf 40(5)}, 762--763. doi: \href{http://dx.doi.org/10.1119/1.1986635}{10.1119/1.1986635}
\bibitem[Deutschman(1977)]{Deutschman1977} Deutschman, W.A. 1977: ``Parallax without pain'' in {\em American Journal of Physics} {\bf 45(5)}, 490. doi: \href{http://dx.doi.org/10.1119/1.11009}{10.1119/1.11009} 
\bibitem[Ferguson(1977)]{Ferguson1977} Ferguson, J.L. 1977: ``More parallax without pain'' in {\em American Journal of Physics} {\bf 45(12)},
1221--1222. doi: \href{http://dx.doi.org/10.1119/1.10697}{10.1119/1.10697}
\bibitem[Grijs(2011)]{Grijs2011} de Grijs, R. 2011: {\em An Introduction to Distance Measurement in Astronomy.} Wiley \& Sons.
\bibitem[Hirshfeld(2013)]{Hirshfeld2013} Hirshfeld, A.W. 2013: {\em Parallax: The Race to Measure the Cosmos.} Dover Publications: Mineola.
\bibitem[van Leeuwen(2007)]{vanLeeuwen2007} van Leeuwen, F. (2007): ``Validation of the new Hipparcos reduction'' in {\em Astronomy and Astrophysics} {\bf 474}, 653--664. doi: \href{http://dx.doi.org/10.1051/0004-6361:20078357}{10.1051/0004-6361:20078357}
\bibitem[Penselin et al.(2014)]{PenselinEtAl2014} Penselin, M., C. Liefke \& M. Metzendorf 2014: ``Zweifacher Blick auf erdnahen Asteroiden'' in {\em Sterne und Weltraum} {\bf 11/2014}, 72--77.
\bibitem[Perryman(2010)]{Perryman2010} Perryman, M. 2010:  {\em The Making of History's Greatest Star Map}. Springer. doi: \href{http://dx.doi.org/10.1007/978-3-642-11602-5\_5}{10.1007/978-3-642-11602-5\_5}
\bibitem[P\"ossel(2017a)]{Poessel2017a} P\"ossel, M. 2017a: ``Parallax: reaching the stars with geometry'' in {\em Science in School} {\bf 39}, 40--44. Online version: \href{http://www.scienceinschool.org/content/parallax-reaching-stars-geometry}{http://www.scienceinschool.org/content/parallax-reaching-stars-geometry}
\bibitem[P\"ossel(2017b)]{Poessel2017b} P\"ossel, M. 2017b: ``Finding the scale of space'' in {\em Science in School} {\bf 40}, 40--45. Online version: \href{http://www.scienceinschool.org/content/finding-scale-space}{http://www.scienceinschool.org/content/finding-scale-space}
\bibitem[Prusti(2012)]{Prusti2012} Prusti, T. 2012: ``The promises of Gaia'' in {\em Astronomische Nachrichten} {\bf 333}, 454--459. doi: \href{http://dx.doi.org/10.1002/asna.201211688}{10.1002/asna.201211688}
\bibitem[Ratcliff et. al.(1993)]{RatcliffEtAl1993} Ratcliff, S.J. et al. 1993: ``The measurement of astronomical parallaxes with CCD imaging cameras on small telescopes'' in {\em American Journal of Physics} {\bf 61}, 208--216. doi: \href{http://dx.doi.org/10.1119/1.17292}{10.1119/1.17292}
\bibitem[Webb(1999)]{Webb1999} Webb, S. 1999: {\em Measuring the Universe: The Cosmological Distance Ladder.} Springer: Berlin, Heidelberg, New York. 
 \end{thebibliography}
\end{document}